\documentclass[12pt]{article}
\begin{document}

\title{Correlation Functions of Multisite Interaction Spin-S models on the
Bethe-like Lattices}
\author{R.G.~Ghulghazaryan\\[1mm]
{\small \sl Department of Theoretical Physics, 
Yerevan Physics Institute,} \\
{\small \sl Alikhanian Br.2, 375036 Yerevan, Armenia}}

\maketitle

\begin{abstract}
Multisite interaction spin-S models in an external magnetic field 
are studied recursively on the Bethe-like lattices. The transfer-matrix method
is extended to calculate exactly the two-spin correlation functions. The
exact expressions for the correlation length and magnetic susceptibility are
derived for spin-1/2 models. The singularity of the correlation length with critical index $\nu
=1$ and the proportionality of magnetic susceptibility to correlation length
in the second order phase transition region of spin-1/2 ferromagnetic models
on the Bethe-like lattices are established analytically.
\end{abstract}

\section{Introduction}
\noindent

The investigation of physical systems such as binary alloys~\cite
{Styer}, classical fluids~\cite{Grim}, liquid bilayers~\cite{Scott}, %
 solid $^3He$~\cite{Roger}, rare gases~\cite{Barker} and anisotropic magnets such as $%
CeBi$, $EuSe$, etc. lead to multisite interaction (MSI) spin and Heisenberg
models. This models have a complex phase diagrams and one of the challenging
problems in modern statistical mechanics is the effect of multisite
interactions on the properties of order-disorder and order-order phase
transitions. Mean-field theory which is most used and the simplest method
for obtaining approximate results, has been shown to give poor results for
such systems~\cite{Roger,Thom,Heringa}. Therefore, to obtain
 better approximations high temperature series expansions~\cite
{Roger}, Monte Carlo simulations~\cite{Heringa}, Bethe lattice and Cayley
tree approximations~\cite{Monroe}-\cite{Thompson} are used. Although,
there are lots of approximate methods one need to have simple but accurate
technique to approximate multisite interaction (MSI) spin systems. After the
pioneer works of Bethe and Peierls~\cite{Bethe} the Bethe lattice
approximation has attracted particular interest because the calculations on
the Bethe lattice are rather simple and results reflects essential features of the systems,
even when conventional mean-field theories fail~\cite{Gujrati}. The main
difference between the Cayley tree and Bethe lattice as discussed in 
Ref. [13] is that in Cayley tree the surface plays a very important
role. The number of sites on it's surface is a finite fraction of the total
sites even in the thermodynamic limit. As a consequence, the models on the
Cayley tree exhibit quite unusual critical behavior without long-range 
order~\cite{Eggarter}-\cite{Yang}. To overcome the problems with boundary one
considers only properties of sites deep within the interior of the Cayley
tree. The union of such equivalent sites is regarded as forming Bethe
lattice and it is assumed to have a translational symmetry like any
regular lattice.

It was shown by Monroe~\cite{Monroe} that one can obtain much better
approximations than on the Bethe lattice by using Husimi or more general
Bethe-like lattices to investigate MSI systems, especially in case of 
frustration. By going to Bethe-like lattice one constructs the system from 
different type of blocks and, hence, can differentiate between two lattices 
having the same coordination numbers, such as triangle and simple cubic 
lattices, whereas in the case of Bethe lattice it is impossible. 
Furthermore, the various options regarding the connection of the basic 
building blocks, the symmetry of Hamiltonian and, as a consequence, the 
possibility of division of such lattices on shells are of 
importance (see Sect. 2).

In the investigation of critical phenomena the correlation functions play an
important role. It is established that near second order phase transition
critical temperature $T_c$ the fluctuations in system rise and grappled all
the system at $T_c$. It is believed that singularities of thermodynamic
quantities such as the specific heat and the susceptibility at $T_c$ are
related to the increase of fluctuations in the system or, in other words, to
divergence of correlation length at $T_c$~\cite{Stanley}. Since two-spin
correlation functions can be measured experimentally as linear response to
an adiabatic or isotermic applied field, or scattering of neutrons or
electromagnetic waves~\cite{Stinchcombe}-\cite{Destri} it is desirable to
have an analytic form of the correlation function to locate phase
transitions and a deeper understanding of phase transition phenomena.

Recently spin-spin correlation functions of spin-$S$ models in an external
magnetic field was calculated on the Bethe lattice~\cite{Izmailian}, where
the crucial role of Bethe lattice dimensionality was demonstrated in
determining the critical behavior of correlation length. It was shown
 that correlation length diverges at the critical point of $m$ and $%
\chi $ for the Bethe lattice Ising model with critical exponents $\nu =1
$, which is different from the mean-field critical exponent $\nu =\frac
{1}{2}.$

In this paper we apply the transfer matrix method to analytical
calculation of the two-spin correlation functions of the multisite
interaction spin-$S$ models in the presence of an external magnetic field on
the general Bethe-like lattices with Hamiltonian symmetric under the
permutation of spins at any $p$-polygon\footnote{$p$-polygon is a closed polygon with $p$ edges
(sites), for example $2$-polygon is a line, $3$-polygon is a triangle etc.}
 of the lattice. The derivation of the analytic %
expression for correlation functions for arbitrary spin-$S$ is presented in 
Sect. 2. For spin-$\frac {1}{2}$ case, taking into account the dimensionality
of the Bethe-like lattice, the exact expressions of the correlation length $%
\xi $ and magnetic susceptibility $\chi $ are derived in Sect. 3. It is demonstrated
exactly that at $T=T_c$ the correlation length $\xi $ diverges with critical
exponent $\nu =1$ and $\chi \sim \xi $ for arbitrary $p$-polygon
Bethe-like lattice with $q$ $p$-polygons going out from each site.

\section{The Model}
\noindent

The Bethe-like lattice may be constructed as follows : Connecting $(q-1)$ $p$%
-polygons at a single base site one gets a first generation branch
(Fig.1.(a)). Connecting $(q-1)$ first generation branches to each site of a
new $p$-polygon (basic block), except the base site of this system one gets the
second generation branch (Fig. 1.(b)). Continuing this process one gets
higher generation branches. To complete the system, one connects $q$ $nth$ %
generation branches at their base site ($0th$ or central site) (Fig.
1.(c)). The set of sites deep within this system is called the Bethe-like
lattice (see the Introduction). As we see only topological and connectivity
properties are used to define the Bethe-like lattice. 
\setlength{\unitlength}{4mm}
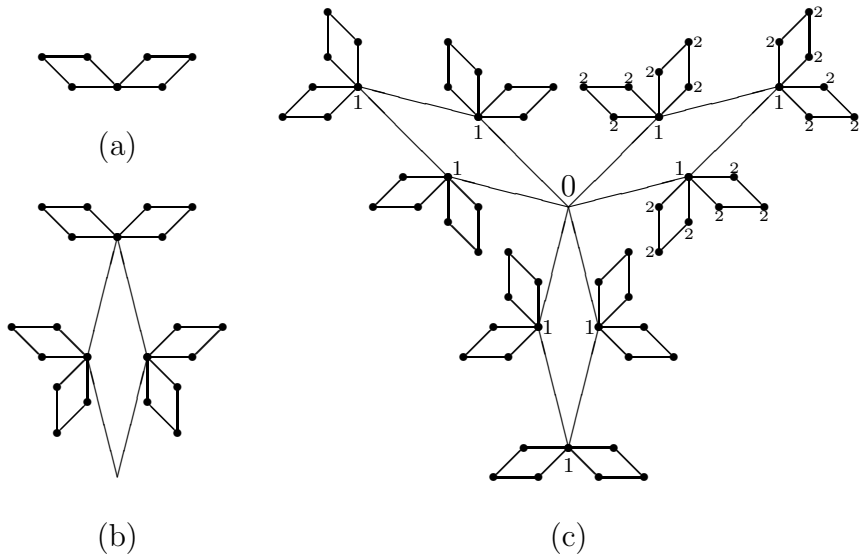
\begin{figure}[htbp]
\vspace*{13pt}
\begin{center}
\newsavebox{\sqa} \savebox{\sqa}(2.5,1){%
\begin{picture}(2.5,1)
\multiput(0,1)(1.5,0){2}{\line(1,-1){1}}
\multiput(0,1)(1,-1){2}{\line(1,0){1.5}}
\put(1,0){\circle*{0.3}}
\put(2.5,0){\circle*{0.3}}
\put(1.5,1){\circle*{0.3}}
\put(0,1){\circle*{0.3}}
\end{picture}}

\newsavebox{\sqb} \savebox{\sqb}(2.5,1){%
\begin{picture}(2.5,1)
\multiput(0,0)(1,1){2}{\line(1,0){1.5}}
\multiput(0,0)(1.5,0){2}{\line(1,1){1}}
\put(0,0){\circle*{0.3}}
\put(1.5,0){\circle*{0.3}}
\put(1,1){\circle*{0.3}}
\put(2.5,1){\circle*{0.3}}
\end{picture}}

\newsavebox{\sqc} \savebox{\sqc}(1,2.5){%
\begin{picture}(1,2.5)
\multiput(0,1)(1,-1){2}{\line(0,1){1.5}}
\multiput(0,1)(0,1.5){2}{\line(1,-1){1}}
\put(0,1){\circle*{0.3}}
\put(1,0){\circle*{0.3}}
\put(1,1.5){\circle*{0.3}}
\put(0,2.5){\circle*{0.3}}
\end{picture}}

\newsavebox{\sqd} \savebox{\sqd}(1,2.5){%
\begin{picture}(1,2.5)
\multiput(0,0)(1,1){2}{\line(0,1){1.5}}
\multiput(0,0)(0,1.5){2}{\line(1,1){1}}
\put(0,0){\circle*{0.3}}
\put(0,1.5){\circle*{0.3}}
\put(1,2.5){\circle*{0.3}}
\put(1,1){\circle*{0.3}}
\end{picture}}

\begin{picture}(30,18)
\put(4,15){\usebox{\sqb}}
\put(1.5,15){\usebox{\sqa}}

\multiput(4,2)(-1,4){2}{\line(1,4){1}}
\multiput(4,2)(1,4){2}{\line(-1,4){1}}
\put(4,10){\usebox{\sqb}}
\put(1.5,10){\usebox{\sqa}}
\put(0.5,6){\usebox{\sqa}}
\put(2,3.5){\usebox{\sqd}}
\put(5,6){\usebox{\sqb}}
\put(5,3.5){\usebox{\sqc}}

\put(9,0){\begin{picture}(20,18)
\multiput(10,3)(-1,4){2}{\line(1,4){1}}
\multiput(10,3)(1,4){2}{\line(-1,4){1}}
\multiput(10,11)(3,3){2}{\line(4,1){4}}
\multiput(10,11)(4,1){2}{\line(1,1){3}}
\multiput(10,11)(-3,3){2}{\line(-4,1){4}}
\multiput(10,11)(-4,1){2}{\line(-1,1){3}}

\put(7.5,2){\usebox{\sqb}}
\put(10,2){\usebox{\sqa}}
\put(6.5,6){\usebox{\sqb}}
\put(8,7){\usebox{\sqc}}
\put(11,6){\usebox{\sqa}}
\put(11,7){\usebox{\sqd}}
\put(13,9.5){\usebox{\sqd}}
\put(14,11){\usebox{\sqa}}
\put(17,14){\usebox{\sqa}}
\put(17,15){\usebox{\sqd}}
\put(13,14){\usebox{\sqd}}
\put(10.5,14){\usebox{\sqa}}
\put(7,14){\usebox{\sqb}}
\put(6,14){\usebox{\sqc}}
\put(2,15){\usebox{\sqc}}
\put(0.5,14){\usebox{\sqb}}
\put(3.5,11){\usebox{\sqb}}
\put(6,9.5){\usebox{\sqc}}

\put(10,0){\makebox(0,0){(c)}}
\put(10,11.7){\makebox(0,0){0}}
\put(13.7,12.3){\makebox(0,0){\scriptsize 1}}
\put(13,13.5){\makebox(0,0){\scriptsize 1}}
\put(17,14.5){\makebox(0,0){\scriptsize 1}}
\put(6.3,12.3){\makebox(0,0){\scriptsize 1}}
\put(7,13.5){\makebox(0,0){\scriptsize 1}}
\put(3,14.5){\makebox(0,0){\scriptsize 1}}
\put(10.7,7){\makebox(0,0){\scriptsize 1}}
\put(9.3,7){\makebox(0,0){\scriptsize 1}}
\put(10,2.4){\makebox(0,0){\scriptsize 1}}

\put(12.7,11){\makebox(0,0){\tiny 2}}
\put(12.7,9.5){\makebox(0,0){\tiny 2}}
\put(14,10.1){\makebox(0,0){\tiny 2}}
\put(15,10.7){\makebox(0,0){\tiny 2}}
\put(16.5,10.7){\makebox(0,0){\tiny 2}}
\put(15.5,12.3){\makebox(0,0){\tiny 2}}
\put(18,13.7){\makebox(0,0){\tiny 2}}
\put(19.5,13.7){\makebox(0,0){\tiny 2}}
\put(18.7,15.3){\makebox(0,0){\tiny 2}}
\put(18.3,16){\makebox(0,0){\tiny 2}}
\put(18.3,17.5){\makebox(0,0){\tiny 2}}
\put(16.7,16.5){\makebox(0,0){\tiny 2}}
\put(14.3,15){\makebox(0,0){\tiny 2}}
\put(14.3,16.5){\makebox(0,0){\tiny 2}}
\put(12.7,15.5){\makebox(0,0){\tiny 2}}
\put(12,15.3){\makebox(0,0){\tiny 2}}
\put(10.5,15.3){\makebox(0,0){\tiny 2}}
\put(11.5,13.7){\makebox(0,0){\tiny 2}}
\end{picture}}

\put(4,13){\makebox(0,0){(a)}}
\put(4,0){\makebox(0,0){(b)}}
%\footnotesize 
\end{picture}
\end{center}

\vspace*{13pt}
\caption{The recursive structure of $4$-polygon Bethe-like lattice with $q=3$.
(a) The first generation branch; (b) the second generation branch; 
(c) the central part of $4$-polygon Bethe-like lattice with $q=3$, 
the numbers $0,1,2$ indicate shells.}
\end{figure}
This system is not a lattice in a conventional sense. The Bethe and
Bethe-like lattices are usually viewed as ramified trees embedded in
infinite dimensional Euclidean space, with a constant vertex connectivity. The fact
that Bethe and Bethe-like lattices usually are viewed as
''infinite-dimensional'' follows from the definition of lattice
dimensionality $d$ in Euclidean space given in Ref. [13] 
\begin{equation}
d=\lim _{n\rightarrow \infty }\frac{\ln \tilde {C}_n}{\ln n},  \label{d}
\end{equation}
where $\tilde{C}_n$ is the total number of sites within $n$ steps of a given site.
For a Bethe-like lattice
\begin{equation}
\tilde{C}_n=1+\sum_{m=1}^{n}C_m,\quad C_n=q(p-1)\left[ (q-1)(p-1)\right] ^{n-1},
\quad C_0=1,  \label{cn}
\end{equation}
where $C_n$ is the number of sites on the nth shell.
The substitution of this expression into (\ref{d}) gives $d=\infty $, so in
this sense the Bethe-like lattice is ``infinite-dimensional''.

Recently it was shown that Bethe-like lattices can be 
embedded in two-dimensional space of constant negative curvature (the H2
hyperbolic plane) with fixed bond angles and lengths. This hierarchical
structures like Bethe-like lattices can be described as a conformal tiling
of the hyperbolic plane and their metric properties are studied~\cite
{Mosseri}.

After this brief survey of the geometrical properties of Bethe-like lattices
let us now define the Hamiltonian of the multisite interaction (MSI) system.
The general MSI spin Hamiltonian have the form 
\begin{equation}
{\cal H}=-\frac{J_p}{S^p}\sum_{\left\langle polygons\right\rangle }\prod S_j-\frac
hS\sum_iS_i,  \label{ham}
\end{equation}
where $S_i$ takes values $S,S-1,\ldots ,-S+1,-S$; the first sum goes over
all the $p$-polygons of the lattice and $\prod $ means the product of all
spins placed on a $p$-polygon; the second term presents the interaction with
an external magnetic field and the sum goes over all the sites of the
lattice.

The advantage of the Bethe-like lattice is that for the models formulated on
it, the exact recurrence relation can be derived and the dynamical systems
theory may be used to explore the thermodynamic properties of the models~\cite
{Eggarter,Thompson,Ghul}. Let us denote the partition function 
of the $N-th$ generation branch with the basic spin in state $S$ as $B(N,S)$.
By cutting the lattice at the basic $p$-polygon one will obtain $(q-1)(p-1)$
interacting $(N-1)-th$ generation branches 
\begin{equation}
B(N,S_0)=\sum_{S_1,\ldots ,S_{p-1}}w(S_0,S_1,\ldots ,S_{p-1})B^\gamma
(N-1,S_1)\cdots B^\gamma (N-1,S_{p-1}),  \label{bns}
\end{equation}
where $S_0,S_1,\ldots ,S_{p-1}$ are the spins at the basic $p$-polygon, $%
w(S_0,S_1,\ldots ,S_{p-1})$ is the statistical weight of the basic p-polygon
and $\gamma =q-1$. 
\begin{equation}
w(S_0,S_1,\ldots ,S_{p-1})=\exp \left\{ \frac{\beta J_p}{S^p}S_0S_1\cdots S_{p-1}+%
\frac{\beta h}{qS}(S_0+\cdots +S_{p-1})\right\},   \label{w}
\end{equation}
where $\beta =1/kT$. By cutting apart the lattice at the central site one
can obtain for the partition function ${\cal Z}=\sum e^{-\beta {\cal H}}$ 
the following expression 
\begin{equation}
{\cal Z}_N=\sum_{S_0}B^q(N,S_0).  \label{z}
\end{equation}
From Eq.~(\ref{w}) follows that $w(S_0,S_1,\ldots ,S_{p-1})$ is a symmetric
function under the permutation of its arguments $S_0,S_1,\ldots ,S_{p-1}$.
Now it is easy to check that all the spins at the $n-th$ generation branch
surface are equivalent and we refer to them as $n-th$ shell (see Fig. 1.(c)).

Let us now define the auxiliary quantities $x(s)$ as follows 
\begin{equation}
x_N(s)=\frac{B(N,s)}{B(N,S)},\;\; x_N(S)=1
{\textstyle\qquad and \qquad} s=S-1,\ldots,-S. \label{x}
\end{equation}
The thermodynamic quantities such as the magnetization, the specific heat
etc. and even thermodynamic potentials such as the free energy\cite
{Ananikian} may be expressed in terms of $x(s)$. Of course, $x(s)$ has no
direct physical meaning, but $\{x(s)\}$ determine the state of the system in
the thermodynamic limit. From Eqs.~(\ref{bns}) and (\ref{x}) we can get $2S$
recurrence equations for $\{x(s)\}$%
\begin{equation}
x_{N+1}(s)= \frac{\sum_{S_1,\ldots ,S_{p-1}}w(s,S_1,
\ldots,S_{p-1})x_N^\gamma (S_1)\cdots x_N^\gamma (S_{p-1})}
{\sum_{S_1,\ldots,S_{p-1}}
w(S,S_1,\ldots ,S_{p-1})x_N^\gamma (S_1)\cdots x_N^\gamma (S_{p-1})}.
\label{xn}
\end{equation}
Using Eqs.~(\ref{z}) and (\ref{x}) one can express the magnetization of the
central site as follows 
\begin{equation}
m_\mu =\left\langle \left( \frac{S_0}S\right) ^\mu \right\rangle =\frac
1{S^\mu }\frac{\sum_{S_0}S_0^\mu x_N^q(S_0)}
{\sum_{S_0}x_N^q(S_0)},  \label{m}
\end{equation}
where $\left\langle \ldots \right\rangle $ denotes the thermal average and $%
\mu $ takes values from $1$ to $2S$.

Further we will restrict our treatment to the case when the series of solutions
of recurrence equations (\ref{xn}) converges to a stable fixed point as $%
N\rightarrow \infty$. Note that for stable fixed point of Eq.~(\ref{xn}) the
difference between $x_N(s)$ and $x_{N-i}(s)$ disappears for $N\rightarrow
\infty $ 
\[
\lim_{N\rightarrow \infty }x_{N-i}(s)=x(s)
\]
for all finite values $i$. The corresponding fixed points equations of the mapping~(%
\ref{xn}) have the form 
\begin{equation}
\sum_{S_i}\left[ w(s,S_1,\ldots,S_{p-1})-x(s)w(S,S_1,\ldots
,S_{p-1})\right] x^\gamma (S_1)\cdots x^\gamma (S_{p-1})=0,  \label{fix}
\end{equation}
where $i=1,\ldots,p-1$.

The correlation functions are defined as follows 
\begin{equation}
g(n)=\frac 1{S^2}\left( \left\langle S_0S_n\right\rangle -\left\langle
S_0\right\rangle \left\langle S_n\right\rangle \right),   \label{gn}
\end{equation}
where spin $S_0$ belongs to the central polygon and $S_n$ located somewhere in the 
$nth$ shell. Using denotation (\ref{x}) and recurrence relations (\ref{bns}) 
we can express $\left\langle S_0S_n\right\rangle $ as follows 
\begin{eqnarray}
\left\langle S_0 S_n\right\rangle  &=&\lim_{N\rightarrow \infty} 
\frac{\sum S_0x_N^\gamma (S_0)w(S_0,y_1,\ldots ,t_1,z_1)x_{N-1}^\gamma
(y_1)\cdots x_{N-1}^\gamma (t_1)\times }{\sum x_N^\gamma
(S_1)w(S_1,y_1,\ldots ,t_1,z_1)x_{N-1}^\gamma (y_1)\cdots x_{N-1}^\gamma
(t_1)\times }  \nonumber \\
&&\hspace{-1cm}
\frac{x_{N-1}^{\gamma -1}(z_1)w(z_1,y_2,\ldots ,t_2,z_2)x_{N-2}^\gamma
(y_2)\cdots x_{N-2}^\gamma (t_2)\times \cdots }{x_{N-1}^{\gamma
-1}(z_1)w(z_1,y_2,\ldots ,t_2,z_2)x_{N-2}^\gamma (y_2)\cdots x_{N-2}^\gamma
(t_2)\times \cdots }  \label{ss} \\
&&\hspace{-1cm}
\frac{x_{N-n+1}^{\gamma -1}(z_{n-1})w(z_{n-1},y_n,\ldots
,t_n,S_n)x_{N-n}^\gamma (y_n)\cdots x_{N-n}^\gamma (t_n)x_{N-n}^\gamma
(S_n)S_n}{x_{N-n+1}^{\gamma -1}(z_{n-1})w(z_{n-1},y_n,\ldots
,t_n,S_n)x_{N-n}^\gamma (y_n)\cdots x_{N-n}^\gamma (t_n)x_{N-n}^\gamma (S_n)},
\nonumber
\end{eqnarray}
where sums run over all spin states displayed.

To use the transfer matrix method let us introduce the following symmetric
matrixes with elements 
\begin{eqnarray}
M_i(z_i,z_{i+1}) &=&\left[ x_{N-i}(z_i)x_{N-i-1}(z_{i+1})\right] ^{\frac{%
\gamma -1}2}\times   \nonumber \\
&&\hspace{-1.5cm}
\sum_{y_{i+1},\ldots ,t_{i+1}} w(z_i,y_{i+1},\ldots
,t_{i+1},z_{i+1})x_{N-i-1}^\gamma (y_{i+1})\cdots x_{N-i-1}^\gamma
(t_{i+1}),\qquad   \label{M} \\
A(S_n,S_0) &=&\left[ x_{N-n}(S_n)x_N(S_0)\right] ^{\frac{\gamma +1}2},
\label{A} \\
A^{\prime }(S_n,S_0) &=&S_nS_0\left[ x_{N-n}(S_n)x_N(S_0)\right] ^{\frac{%
\gamma +1}2}.  \label{AA}
\end{eqnarray}
For stable fixed points $\{x(s)\}$ of Eq.~(\ref{xn}) all matrixes $M_i$ 
coincide, $M_i\equiv M$ for all $i$. Thus, we can rewrite $\left\langle
S_0S_n\right\rangle $ as follows 
\begin{equation}
\left\langle S_0S_n\right\rangle =\frac{Tr(A^{\prime }M^n)}{Tr(AM^n)}.
\label{tr}
\end{equation}
Let us now briefly discuss some properties of $M_i$ matrixes. Above we stated
that for stable fixed points $\{x(s)\}$ all the matrixes $M_i$ coincide,
but this is true only when $w(S_0,S_1,\ldots ,S_{p-1})$ is a symmetric
function under the permutations of its arguments, which holds for MSI
Hamiltonian (\ref{ham}). In the case when one wants to introduce also pair
interactions one find that the definition of $M_i$ becomes not unique. To
demonstrate this let us consider spin-$\frac 12$ Ising model on the $4$%
-polygon (square) Bethe-like lattice with $J_2$ pair interactions only. In
Fig. 2 one can see two configurations of spins on a $4$-polygon which
differ only in permutation of spins, but their statistical weights are
different. For example, for configuration (1) we have 
\begin{eqnarray*}
M(+,+) &=& \sum_{y_{i+1},\ldots ,t_{i+1}}
w(+,y_{i+1},t_{i+1},+)x^\gamma (y_{i+1})x^\gamma (t_{i+1})= \\
&&e^{4\beta J_2}+x^{2\gamma }+2x^\gamma 
\end{eqnarray*}
and for configuration (2) respectively 
\[
M(+,+)=e^{4\beta J_2}+e^{-4\beta J_2}x^{2\gamma }+2x^\gamma. 
\]

\begin{figure}[htbp]
\vspace*{13pt}

\begin{center}
\newsavebox{\weg} \savebox{\weg}(2.5,2.5){%
\begin{picture}(2.5,2.5)

\multiput(1,1)(0,4){2}{\line(1,0){4}}
\multiput(1,1)(4,0){2}{\line(0,1){4}}

\put(1,1){\circle*{0.3}}
\put(1,5){\circle*{0.3}}
\put(5,5){\circle*{0.3}}
\put(5,1){\circle*{0.3}}
\end{picture}}

\begin{picture}(20,7)
\put(2,1){\usebox{\weg}}
\put(5,0){\makebox(0,0){$w(+,-,-,+)=1$}}
\put(13,1){\usebox{\weg}}
\put(16,0){\makebox(0,0){$w(+,-,-,+)=e^{-4\beta J_2}$}}
\put(4.5,4){(1)}
\put(15.5,4){(2)}

\put(2,2){$+$}
\put(2,6){$+$}
\put(7.3,6){$-$}
\put(7.3,2){$-$}

\put(13,2){$+$}
\put(13,6){$-$}
\put(18.3,6){$+$}
\put(18.3,2){$-$}

\end{picture}
\end{center}

\vspace*{13pt}
\caption{Two configuration of spins on a $4$-polygons with asymmetric 
statistical weight function.}
\end{figure}
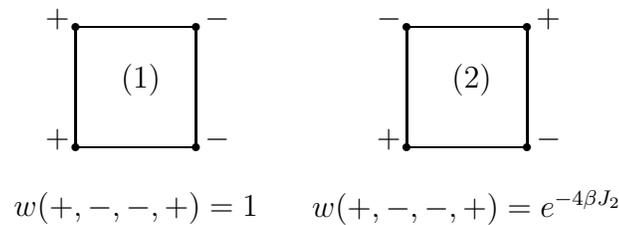

To recover the symmetry of $w$ we can add all possible diagonal interactions
between the spins on a polygon (see Fig. 3). Thus, if one wants to consider
pair or /and $l$-interactions ($l=2,\ldots,p-1$ is the number of interacting 
spins on a $p$-polygon) one has to consider all possible interactions between
spins on a polygon to have a symmetric function of the statistical weight $w$
under the permutation of spins on a $p$-polygon.

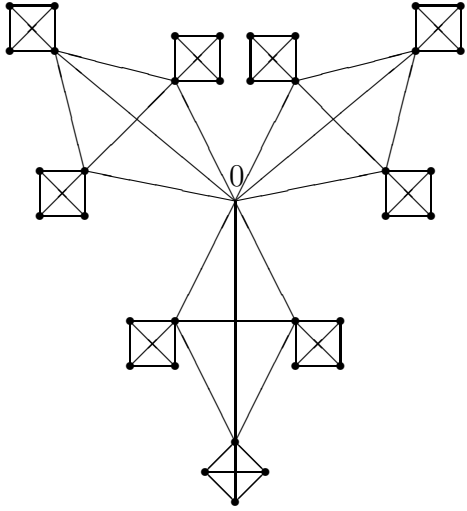
\begin{figure}[htbp]
\vspace*{13pt}

\begin{center}
\newsavebox{\sq} \savebox{\sq}(2.5,2.5){%
\begin{picture}(2.5,2.5)
\multiput(0,0)(0,1.5){2}{\line(1,0){1.5}}
\multiput(0,0)(1.5,0){2}{\line(0,1){1.5}}
\put(0,0){\line(1,1){1.5}}
\put(1.5,0){\line(-1,1){1.5}}
\put(0,0){\circle*{0.3}}
\put(1.5,0){\circle*{0.3}}
\put(1.5,1.5){\circle*{0.3}}
\put(0,1.5){\circle*{0.3}}
\end{picture}}

\newsavebox{\sqq} \savebox{\sqq}(2,2){%
\begin{picture}(2,2)
\multiput(1,0)(-1,1){2}{\line(1,1){1}}
\multiput(1,0)(1,1){2}{\line(-1,1){1}}
\put(1,0){\line(0,1){2}}
\put(0,1){\line(1,0){2}}
\put(1,0){\circle*{0.3}}
\put(0,1){\circle*{0.3}}
\put(1,2){\circle*{0.3}}
\put(2,1){\circle*{0.3}}
\end{picture}}

\begin{picture}(18,17)

\multiput(9,3)(-2,4){2}{\line(1,2){2}}
\multiput(9,3)(2,4){2}{\line(-1,2){2}}
\put(9,3){\line(0,1){8}}
\put(7,7){\line(1,0){4}}

\put(9,11){\line(5,1){5}}
\put(9,11){\line(6,5){6}}
\put(9,11){\line(1,2){2}}
\put(11,15){\line(4,1){4}}
\put(11,15){\line(1,-1){3}}
\put(14,12){\line(1,4){1}}

\put(9,11){\line(-5,1){5}}
\put(9,11){\line(-6,5){6}}
\put(9,11){\line(-1,2){2}}
\put(4,12){\line(-1,4){1}}
\put(4,12){\line(1,1){3}}
\put(7,15){\line(-4,1){4}}

\put(8.8,11.5){0}

\put(11,5.5){\usebox{\sq}}
\put(5.5,5.5){\usebox{\sq}}
\put(14,10.5){\usebox{\sq}}
\put(2.5,10.5){\usebox{\sq}}
\put(15,16){\usebox{\sq}}
\put(1.5,16){\usebox{\sq}}
\put(9.5,15){\usebox{\sq}}
\put(7,15){\usebox{\sq}}
\put(8,1){\usebox{\sqq}}

\end{picture}
\end{center}
\vspace*{13pt}
\caption{The schematic of the $4$-polygon Bethe-like lattice with all possible 
pair interactions on a polygon.}
\end{figure}

In order to use the transfer-matrix method we consider only the case when $w$
is a symmetric function as described above. Now let us return to matrices $M$
and $A$. Using the symmetry property of $w$, Eq.~(\ref{bns}) and definitions (%
\ref{M}), (\ref{A}) one can easily prove that matrices $A$ and $M$ 
commute. Thus they can be diagonalized simultaneously. The spectrum of
matrix $A$ is highly degenerated, it has only one nonzero eigenvalue 
\[
\lambda _1^A=1+\sum_{s=S-1,\ldots,-S}x^{\gamma +1}(s)
\]
with eigenvector 
\[
v_1^\ast=\left( \lambda _1^A\right) ^{-\frac 12}\left( 1,x^{\frac{\gamma +1}%
2}(S-1),\ldots ,x^{\frac{\gamma +1}2}(-S)\right) .
\]
After the diagonalization of matrix $M$ we come to the following
expression for the correlation functions 
\begin{equation}
g(n)=\sum_{k=1}^{2S}A_k\left( \frac{\lambda _{k+1}}{\lambda _1}\right) ^n,
\label{g}
\end{equation}
where $\lambda _k$ are the eigenvalues of matrix $M$. The
coefficients $A_k$ $(k=1,2,\ldots ,2S)$ can be obtained by solving the
following system of linear algebraic equations 
\[ 
\left\{\begin{array}{lll}
g(0) = A_1+A_2+\cdots +A_{2S}, \\
g(1) = A_1l_1+A_2l_2+\cdots +A_{2S}l_{2S}, \\
\vdots  \\
g(2S-1) = A_1l_1^{2S-1}+A_2l_2^{2S-1}+\cdots +A_{2S}l_{2S}^{2S-1},
\end{array}
\right.
\]
where $l_k\equiv \lambda _{k+1}/\lambda _1$. The solution of this system 
of equations can be presented in the form
\begin{equation}
A_k=\frac{\sum_{j=0}^{2S-1} (-1)^{2S-1-j}F_j(k)g(j)}
{\prod_{i=1}^{2S}(l_k-l_i)},\qquad i\neq k.
\label{aka}
\end{equation}
where $F_j(k)$ are the elementary functions of $2S-1$ variables 
$l_1,l_2,\ldots ,l_{k-1},l_{k+1},\ldots ,l_{2S}$:
\[
\begin{array}{lll}
F_{2S-1}(k) =1, \\
F_{2S-2}(k) = l_1+\cdots l_{k-1}+l_{k+1}+\cdots +l_{2S}, \\
\vdots  \\
F_0(k) = l_1\cdots l_{k-1}l_{k+1}\cdots l_{2S}
\end{array}
\]
(for more details see the Appendix of Ref.[21]).
From the Eq.~(\ref{aka}) we can find for MSI spin-$\frac 12$ Ising model 
$\langle {S_0^2}\rangle =\frac 14$ and 
\begin{equation}
A_1=g(0)=4\left\langle S_0^2\right\rangle -m_1^2=1-m_1^2,
\label{a1}
\end{equation}
for MSI spin-$1$ Ising model respectively 
\[
A_1=\frac {l_2g(0)-g(1)}{l_2-l_1}=\frac{\lambda_3(m_2-m_1^2)-\lambda_1
(\left\langle S_0 S_1\right\rangle-m_1^2)}{\lambda_3-\lambda_2}
\]
and $A_2=A_1(l_1\Leftrightarrow l_2)$, where 
$m_2=\left\langle S_0^2\right\rangle$, etc.

Thus, Eqs.~(\ref{m}),(\ref{fix}) and (\ref{g}) give us a full set of
equations for investigation of spin-$S$ MSI and pair interaction models on
the Bethe-like lattices.

\section{The General MSI Spin-$\frac {1}{2}$ Ising Model}
\noindent

Here we apply the general technique developed in previous section to
investigate the MSI spin-$\frac 12$ Ising model in an external magnetic
field $h$ with Hamiltonian 
\begin{equation}
{\cal H}=-2^pJ_p\sum_{<polygons>}\prod S_p-4J_2\sum_{<ij>}S_iS_j-2h\sum_iS_i,
\label{h2}
\end{equation}
where $S_i$ takes values $\pm \frac 12,$ the first term describes the
multisite interactions on p-polygons, the second sum goes over all possible
pairs of spins on each $p$-polygon (see Sect. 2) and the third one goes over
all spins on a lattice. The statistical weight of a $p$-polygon has the
form 
\begin{equation}
w(S_0,\ldots ,S_{p-1})=\exp \left[ 2^p\beta J_pS_0\cdots S_{p-1}+4\beta
J_2\sum_{<ij>\in P}S_iS_j+\frac{2\beta h}q\sum_{i\in P}S_i\right]. 
\label{w2}
\end{equation}
We see that $w$ is a symmetric function under the permutation of its
arguments. Let us make the following notations 
\begin{eqnarray}
a_k &=& w(\;\underbrace{+,\ldots,+}_k,\underbrace{-,\ldots,-}_{p-k}\;)
 \label{ak} \\
&&\textstyle{or}  \nonumber \\
a_k &=& \exp \left[ (-1)^{p-k}\beta J_p+\frac{\left( 2k-p\right) ^2-p}%
2\beta J_2+\frac{\left( 2k-p\right) }q\beta h\right],   \nonumber
\end{eqnarray}
where $k$ is the number of spins on a $p$-polygon in the state $+\frac 12,$ $%
k\leq p$. Using this notations and $x_N(-)\equiv x_N$ we can present the
corresponding recurrence equations (\ref{xn}) as follows 
\begin{equation}
x_N=f(x_{N-1}),\qquad f(x)=\frac{\sum_{k=0}^{p-1}
a_kC_{p-1}^kx^{(p-1-k)\gamma }}
{\sum_{k=0}^{p-1}a_{k+1}C_{p-1}^kx^{(p-1-k)\gamma }},  \label{xn2}
\end{equation}
where $C_{p}^k=p!/\left(k!(p-k)!\right)$ is the number of ways of choosing $k$
elements from $p$ ones without permutations.

The stable fixed points $x$ of the mapping (\ref{xn2}) can be obtained from
the fixed point equation 
\begin{equation}
x=f(x),\qquad \sum_{k=0}^{p-1}\left( a_k-xa_{k+1}\right)
C_{p-1}^kx^{(p-1-k)\gamma }=0  \label{fix2}
\end{equation}
and the stability condition 
\[
\left| \frac \partial {\partial x}f(x)|_{x=x_{c}}\right| <1.
\]
Using the identity $C_{p-1}^k=C_{p-2}^k+C_{p-2}^{k-1}$ one can rewrite 
the fixed point equation (\ref{fix2}) as follows 
\begin{equation}
\sum_{k=0}^{p-2}\left( x^{\gamma +1}a_{k+1}-x^\gamma
a_k+xa_{k+2}-a_{k+1}\right) C_{p-2}^kx^{(p-2-k)\gamma }=0.  \label{fix22}
\end{equation}
In the stable fixed point the magnetization of the central site has the form 
\begin{equation}
m_1=\frac{1-x^q}{1+x^q}.  \label{m2}
\end{equation}
It is well known that a ferromagnetic system at high temperatures ($T>T_c$)
 is in the %
paramagnetic phase with $m_1=0$ and, hence, x=1 is a stable fixed point of
this phase (see Eqs.~(\ref{fix2})-(\ref{m2})). For low temperatures ($T<T_c$)
the system is in the ferromagnetic phase with $m_1\neq 0$ and, hence, $x\neq
1$. Thus, at the critical point ($T=T_c$) the fixed point $x_c=1$ loses its
stability 
\begin{equation}
\frac \partial {\partial x}f(x)\mid_{x_c=1}=1.  \label{stf}
\end{equation}
Here we demand that coefficients $a_k$ should be taken such that Eq.(\ref{fix2}) 
has $x=1$ solution. 
For antiferromagnetics the critical temperature can be found from  
Eq.~(\ref{fix2}) and the following condition 
\begin{equation}
\frac \partial {\partial x}f(x)\mid_{x=x_c}=-1.
\end{equation}
From definitions (\ref{M})-(\ref{AA}) for stable fixed points labeling rows
(columns) of the matrixes $A$, $A^{\prime }$ and $M$ as $+\frac 12,-\frac 12$
from left to right (from up to down) we have 
\[
M=\left( 
\begin{array}{ll}
\sum_{k=0}^{p-2}a_{k+2}C_{p-2}^kx^{(p-2-k)\gamma } & x^{\frac{\gamma -1}%
2}\sum_{k=0}^{p-2}a_{k+1}C_{p-2}^kx^{(p-2-k)\gamma } \vspace{3mm}\\ 
x^{\frac{\gamma -1}2}\sum_{k=0}^{p-2}a_{k+1}C_{p-2}^kx^{(p-2-k)\gamma } & 
x^{\gamma-1}\sum_{k=0}^{p-2}a_kC_{p-2}^kx^{(p-2-k)\gamma }
\end{array}
\right), 
\]
\begin{equation}
A=\left( 
\begin{array}{ll}
1 & x^{\frac{\gamma +1}2} \\ 
x^{\frac{\gamma +1}2} & x^{\gamma +1}
\end{array}
\right) \textstyle{\qquad and\qquad} A^{\prime }=\left( 
\begin{array}{ll}
1 & -x^{\frac{\gamma +1}2} \\ 
-x^{\frac{\gamma +1}2} & x^{\gamma +1}
\end{array}
\right).   \label{maa}
\end{equation}
Now from Eqs.~(\ref{g}) for two-spin correlation functions we can write 
\begin{equation}
g(n)=A_1\lambda ^n,  \label{g2}
\end{equation}
where $A_1=1-m_1^2$ (see Eq.~\ref{a1}), $\lambda =\lambda _2/\lambda _1$ and $%
\lambda _1,$ $\lambda _2$ are the eigenvalues of the matrix $M$%
\begin{eqnarray}
\lambda _1 &=&\sum_{k=0}^{p-2}(a_{k+2}+x^\gamma
a_{k+1})C_{p-2}^kx^{(p-2-k)\gamma },  \label{lam1} \\
\lambda _2 &=&x^{\gamma-1}\sum_{k=0}^{p-2}(a_k-xa_{k+1})C_{p-2}^kx^{(p-2-k)\gamma }.
\label{lam2}
\end{eqnarray}
Let us now consider the behavior of $g(n)$ in the critical region for the
ferromagnetic models. If one restricts the interactions to pair
interactions ($J_2+h$), the Lee-Yang circle theorem\cite{Lee} establishes
that only at $h=0$ a phase transition can occur. However, with 
multisite interactions present (MSI$+h$), phase transition may occur for 
$h\neq 0$ at $T_c\neq 0$ and for $h=0$ only at $T_c=0$~\cite{Monroe}.

The $T-h$ second order phase transition region ( \textit{surface} for (MSI$%
+J_2+h$), \textit{line} for (MSI$+h$), and a \textit{point} for ($J_2+h$)
interactions) can be obtained from Eq.~(\ref{stf}) 
\begin{equation}
(p-1)(q-1)\sum_{k=0}^{p-2}(a_k-a_{k+1})C_{p-2}^k=%
\sum_{k=0}^{p-1}a_{k+1}C_{p-1}^k.  \label{reg}
\end{equation}
For example, the critical point for ($J_2+h$)
interaction model on the Bethe lattice ($p=2$) is\cite{Baxter}
\[
\left( J_2\right) _c=\frac 12\ln \frac q{q-2}
\]
and for Husimi lattice ($p=3$) is
\[
\left( J_2\right) _c=\frac 14\ln \frac{2q+1}{2q-3}.
\]
Let us now calculate the correlation length function $\xi $. Substituting $%
x_c=1$ into Eqs.~(\ref{g2})-(\ref{lam2}) and taking into account Eq.~(\ref{reg})
for correlation function at the critical region $g_c(n)$ we obtain 
\begin{equation}
g_c(n)=\left[ (p-1)(q-1)\right] ^{-n}.  \label{gc}
\end{equation}
From the definition of the dimensionality presented in Sect. 1 follows that
for $n\gg1$ 
\begin{equation}
d=\frac n{\ln n}\ln \left[ (p-1)(q-1)\right].   \label{dim}
\end{equation}
Hence, $g_c(n)$ can be written as 
\begin{equation}
g_c(n)\sim n^{-d}.  \label{gd}
\end{equation}
Thus, near the critical region the correlation function has the form 
\begin{equation}
g(n)=n^{-d}\exp \left( -\frac n\xi \right),   \label{gksi}
\end{equation}
where the correlation length $\xi $ is
\begin{equation}
\xi =\left[ \ln \left[ \frac 1{(p-1)(q-1)\lambda }\right] \right] ^{-1}.
\label{len}
\end{equation}
The obtained formulas (\ref{gd})-(\ref{len}) are in good agreement with the
general behavior of correlation functions described in Ref. [13].
From Eqs.~(\ref{lam1}), (\ref{lam2}) and (\ref{len}) follows that near the
critical region the correlation length $\xi $ has singularity in the form 
\begin{equation}
\xi \sim \xi (J_p,J_2,h)|_{t=0}\,t^{-1},  \label{sing}
\end{equation}
where 
\[
\xi^{-1} (J_p,J_2,h)=\frac{\sum_{k=0}^{p-2}(a_{k+2}^{\prime }
+a_{k+1}^{\prime}-(p-1)(q-1)(a_k^{\prime }-a_{k+1}^{\prime }))C_{p-2}^k}
{\sum_{k=0}^{p-2}(a_{k+2}+a_{k+1})C_{p-2}^k}
\]
and $t=\frac{T-T_c}{T_c}$ , $a_k^{\prime }=\frac \partial {\partial
t}a_k|_{t=0}$. Thus, the correlation length $\xi $ of ferromagnetic models
on the Bethe-like lattices increases as the critical region is approached
according to $\xi \sim t^{-\nu}$ with critical exponent $\nu =1$.

Further, let us calculate the bulk susceptibility per lattice site $\chi =%
\frac{\partial m_1}{\partial h}$. In our case, because we have the exact
expression for correlation functions, it is easier to calculate $\chi $
using the fluctuation relation\cite{Baxter} 
\begin{equation}
\chi =\left( N_lkT\right) ^{-1}\sum_i\sum_jg_{ij},  \label{xi}
\end{equation}
where the summation goes over all spins, $g_{ij}$ is the correlation
functions of spins on sites $i,j$ and $N_l$ is the total number of
sites on a lattice. Let us now consider the Eq.~(\ref{xi}) in detail 
\begin{equation}
\chi =\frac 1{kT}\lim_{N_{l}\rightarrow \infty }\frac 1{N_l}\sum_i\sum_j\frac
1{S^2}\left( \langle S_iS_j \rangle-\langle S_i\rangle\langle S_j\rangle
\right).
\label{xi2}
\end{equation}
Taking into account the translational invariance and equivalence of all sites
of Bethe-like lattice (see Sect. 2) we have 
\begin{eqnarray*}
\langle S_i\rangle &=& \langle S_j\rangle, \\
\frac 1{N_l}\sum_i\sum_j \langle S_iS_j\rangle &=& \sum_j\langle S_0 S_j\rangle.
\end{eqnarray*}
Substituting this relation into Eq.~(\ref{xi2}) we obtain 
\[
\chi =\frac 1{kT}\lim_{N_l\rightarrow \infty }\sum_j\frac 1{S^2}\left(
\langle S_0S_j\rangle-\langle S_0\rangle^2\right) =
\beta\lim_{l\rightarrow \infty }\sum_{n=0}^lC_ng(n),
\]
where $C_n$ is the number of sites on the $nth$-shell and $l$ is the number
of shells on a lattice. From Eqs.~(\ref{cn}) and (\ref{g2}) one can easily
find 
\begin{eqnarray*}
\chi  &=&\beta (1-m_1^2)\lim_{l\rightarrow \infty }\left[ 1+q(p-1)\lambda
\sum_{n=1}^l\left[ (p-1)(q-1)\lambda \right] ^{n-1}\right] = \\
&&\beta (1-m_1^2)\lim_{l\rightarrow \infty }\left[ \frac{1+\lambda(p-1) }{%
1-(p-1)(q-1)\lambda }-q(p-1)\lambda \frac{\left[ (p-1)(q-1)\lambda \right] ^l%
}{1-(p-1)(q-1)\lambda }\right]. 
\end{eqnarray*}
It is now obvious that for $(p-1)(q-1)\lambda \geq 1$ the susceptibility $%
\chi $ diverges and for $(p-1)(q-1)\lambda \leq 1$ have the form 
\begin{equation}
\chi =\frac{\beta (1-m_1^2)\left( 1+\lambda(p-1) \right) }{1-(p-1)(q-1)\lambda }.
\label{xml}
\end{equation}
From Eqs.~(\ref{sing}) and (\ref{xml}) for the general ferromagnetic MSI 
spin-$\frac 12$ %
model in an external magnetic field one can easily establish the relation
between the susceptibility $\chi $ and the correlation length $\xi $ in the
critical region $\chi \sim \xi $.

\section{Concluding Remarks}
\noindent 
In this paper we have applied the general Bethe-like lattice approximation
to studying the correlation functions of the most general multisite
interaction (MSI) spin-$S$ Ising models in an external magnetic field.
Deriving the recurrence relations and using dynamical systems theory the
thermodynamic properties of the system are studied by the investigation of the
behavior of the fixed points of the corresponding map. Exact formulas for
the correlation functions $g(n)$ (spin-$S$ Ising models), and for
the correlation length $\xi $, the bulk
susceptibility per lattice site $\chi $ (spin-$\frac 12$ Ising models) %
have been derived for arbitrary $h$ %
and $T$. A singular behavior of the correlation length $\xi $ near the
critical region with critical exponent $\nu =1$ has been obtained exactly.
We also showed that near the critical region the magnetic susceptibility $%
\chi $ is proportional to the correlation length $\xi $.

From Eq.~(\ref{gc}) one can see that in the critical region the correlation
functions tend to zero as $n\rightarrow \infty $ and one may try to interpret
this as absence of correlation on the Bethe-like lattice models. But this is
not correct, because as we can see from Eqs.~(\ref{dim}) and (\ref{gd}) the 
fact that at $T=T_c$ $g(n)\rightarrow 0$ is a consequence of infinite
dimensionality of the Bethe-like lattices embedded in the Euclidean space.
As we can see from Eq.~(\ref{sing}) correlation length in fact has singularity
at $T=T_c.$ Hence, the second order phase transitions of models defined on
the Bethe-like lattices have the same ``attributes'' as the models on
regular lattices. As showed by
Gujrati~\cite{Gujrati} Bethe and Bethe-like lattice approximations give
qualitatively correct results even when conventional mean-field theories 
fail. Furthermore, due to various options regarding the connection of the
basic building blocks, using Bethe-like lattices one can get better
approximations with respect to mean-field and usual Bethe lattice
calculations, especially in the presence of frustrations~\cite{Monroe}.

In conclusion we want to note that the method presented here is quite
general and may be used for other models also, for example the Potts model,
the multilayer Ising model etc.

\section{Acknowledgements}
\noindent

Author would like to thank N. S. Ananikyan and N. Sh. Izmailyan for
fruitful discussions. This work was partly supported by the Grants
INTAS-96-690, INTAS-97-347.

\end{document}